%% file: papcp.tex
\documentclass[aps,prd,showpacs,superscriptaddress,reprint]{revtex4-1}
\usepackage{amsmath,amsfonts,amssymb}
\usepackage{dsfont,graphicx}
\usepackage{bm}
\graphicspath{{Plots/}}
\newcommand{\bs}{\begin{subequations}}
\newcommand{\es}{\end{subequations}}
\newcommand{\adb}{\allowdisplaybreaks } 
\newcommand{\ann}{\adb \nonumber \\}
\newcommand{\oa}{\overline{a}}
\newcommand{\TE}{\textsf{TE}}
\newcommand{\TM}{\textsf{TM}}

\newcommand{\cN}{\mathcal{N}}

\usepackage{color} \definecolor{darkgreen}{rgb}{0,.5,0}
\usepackage[colorlinks,filecolor=blue,citecolor=darkgreen,unicode]{hyperref}
\begin{document}
\title{The Casimir-Polder effect for a stack of conductive planes}

\author{Nail Khusnutdinov}\email{nail.khusnutdinov@ufabc.edu.br}
\affiliation{Centro de Matem\'atica, Computa\c{c}\~ao e Cogni\c{c}\~ao, Universidade Federal do ABC, 09210-170 Santo Andr\'e, SP, Brazil}
\affiliation{Institute of Physics, Kazan Federal University, Kremlevskaya 18, Kazan, 420008, Russia}
\author{Rashid Kashapov}\email{kashapov.rashid@gmail.com}
\affiliation{Institute of Physics, Kazan Federal University, Kremlevskaya 18, Kazan, 420008, Russia}
\author{Lilia M. Woods}
\affiliation{Department of Physics, University of South Florida, Tampa, Florida 33620, USA}
\date{\today}
\begin{abstract}
The Casimir-Polder interaction between an atom and a multilayered system composed of infinitely thin planes is considered using the zeta-function regularization approach with summation of the zero-point energies. As a prototype material, each plane is represented by a graphene sheet whose optical response is described by a constant conductivity or Drude-Lorentz model conductivity. Asymptotic expressions for various separations are derived and compared to numerical calculations. We distinguish between large atom/plane limit, where retardation effects are prominent, and small atom/plane limit, where the typical van der Waals coefficient is found to be dependent on the number of graphenes and characteristic distances. The calculated energies for different atoms and graphene conductivity models brings forward the basic science of the Casimir-Polder effect and suggests ways to manipulate this interaction experimentally.

\end{abstract}  
\maketitle 

\section{Introduction}

Interactions originating from electromagnetic fluctuations between objects are of much interest from a fundamental point of view as well as for the development of novel devices. Van der Waals (vdW), Casimir, and Casimir-Polder forces are examples of such interactions. Their common origin has been recognized in the early works by Lifshitz and collaborators \cite{Lifshitz:1956:Ttmafbs,*Lifshitz:1980:SP} and they have been studied extensively in recent years \cite{Klimchitskaya:2009:TCfbrmet,*Dalvit:2011:CP,*Woods:2015:AMPoCavdWI} to advance our understanding of light-matter interactions.  The vdW regime corresponds to small distance separation between the objects, where the speed of light $c$ is neglected. The Casimir force, describing interactions between objects with macrodimensions, and the Casimir-Polder force, describing interactions between polarizable particles and objects with macrodimensions, on the other hand, correspond to the retarded regime.  

The Casimir-Polder force is of great relevance to novel phenomena such as trapping cold atoms near surfaces, Bose-Einstein condensates, and quantum reflection \cite{Leanhardt:2003:BCnaMS,*Lin:2004:IotCPaJNoBCSNS,Druzhinina:2003:EOoQRffT,*Friedrich:2002:QrbCdWpt}. As trapped atoms appear to be very sensitive to the electromagnetic characteristics of the nearby objects, they represent a powerful tool to gain insight and ultimately control the atom-wall coupling. Therefore, the types of materials used for the wall composition can significantly influence the Casimir-Polder force. Recent studies have shown that systems involving graphene  give new perspectives into this problem due to the graphene reduced dimensionality and response properties determined by the Dirac-like energy band structure. Several reports have focused on theoretical calculations of atom/graphene Casimir-Polder interactions. These investigations typically use the Lifshitz theory, which expresses the energy in terms of the atomic polarizability and frequency-dependent response properties of the material. It was shown that the dielectric function of graphene, described via the Dirac and hydrodynamics models, leads to different magnitudes of the Casimir-Polder energy \cite{Churkin:2010:CohaDmodibgaHHoNa}. The much reduced interaction captured via the Dirac model as compared to typical metallic surfaces, has also been suggested as means to shield vacuum Casimir-Polder fluctuations \cite{Ribeiro:2013:Svfwg}. Casimir-Polder thermal effects involving graphene have also been studied showing unusual distance asymptotics when compared to atom/metal wall interactions \cite{Klimchitskaya:2014:Iogcotai,*Chaichian:2012:TCiodawg,*Klimchitskaya:2014:CCfbpmatfig,*Kaur:2014:EtdioLNKaRaawgitDm}. Exploring the extraordinary magneto-optical response of graphene, on the other hand, was suggested as means to control the Casimir-Polder interaction by an applied magnetic field \cite{Cysne:2014:TtCivmeig}.   

Although atom/single graphene and atom/graphene covered substrates have been considered by several authors, as discussed above, the Casimir-Polder interaction in atom/multilayered graphenes is yet to be explored. Previous studies have shown that the Casimir energy in multilayered systems with planar, cylindrical, or spherical symmetries can significantly affect not only the strength, but also the characteristic distance dependences of the interaction  \cite{Ninham:1970:vdWIiMS,*Ninham:1970:vdWFaTF,*Parsegian:2006:VdWFHBCEP,*Zhou:1995:vdWarCioaeoaawmw,*Sernelius:2014:EnmaCeils,*Tatur:2008:ZeNpcccs}. Similarly, the effects of number of graphenes, graphene-graphene separations, and atom-graphene distance are factors that will affect the Casimir-Polder force. In addition, investigating different models of describing the response properties of each graphene will also influence the Casimir-Polder force and lead to distinct asymptotic relations. 

In this paper, we consider the Casimir-Polder interaction in atom/multilayered systems using the zeta-regularization approach, which relies on the the zero-point mode summation of the electromagnetic field. This technique was used to study the Casimir energy in an infinitely thin spherical shell, two graphene sheets, and multi-layered graphenes \cite{Khusnutdinov:2014:Cefswcc,*Khusnutdinov:2015:Cefasocp,Dalvit:2011:CP,*Khusnutdinov:2015:Cefacopcs}. As a prototype to the multilayers we take graphene sheets, where each graphene can be taken into account via its conductivity $\sigma(\omega)$, described with different models. Here we utilize $\sigma$ given by a constant conductivity and Drude-Lorentz approximations. Using media rarefication to obtain the atom/substrate coupling, analytical expressions in various asymptotic limits are obtained, which enable obtaining factors that determine the interaction.

\section{The Casimir-Polder energy}\label{Sec:CPForce}

\begin{figure}[htb]
	\centering
	\includegraphics[width=6truecm]{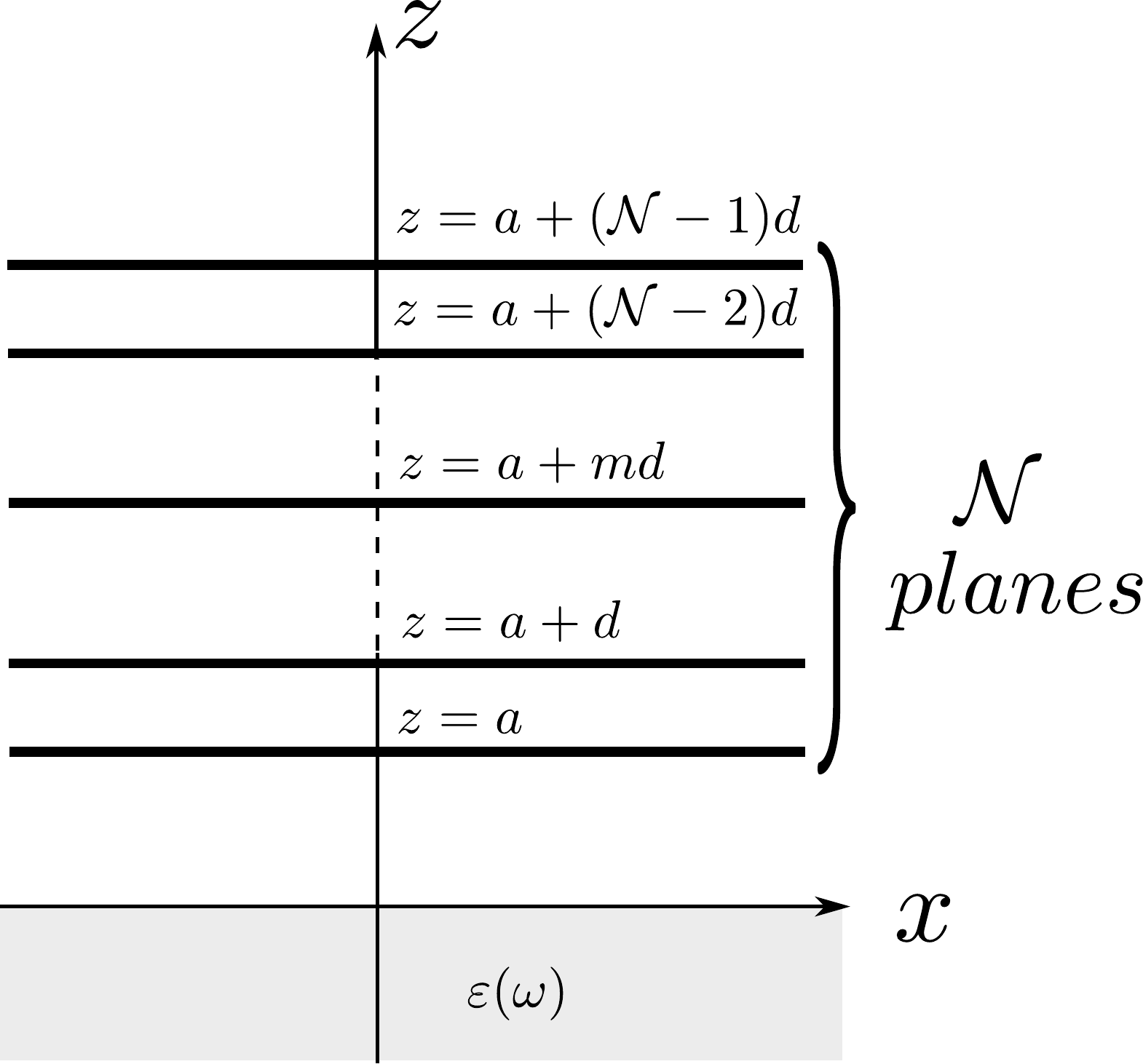}
	\caption{The $\cN$ parallel planes are located at points $z=a,a+d,a+2d,a+3d,\ldots,a+(\cN-1)d$. The half-space $z\leq 0$ is filled by dielectric media with permeability $\varepsilon (\omega)$.}\label{fig:w1}
	\end{figure}

The zeta-regularization technique relies on finding the zero-point energy as a regularized quantity for a given configuration using the appropriate electromagnetic boundary conditions, as outlined in \cite{Bordag:2009:ACE}.  The system under consideration here consists of half of the space $z<0$ being occupied by a material with dielectric function $\varepsilon (\omega)$ and a stack of $\cN$ equally spaced infinitely thin layers positioned above it, as shown in Fig. \ref{fig:w1}. We distinguish between the plane-plane separation $d$ and the substrate-bottom plane separation $a$. The zero-point energy can  be expressed as 
\begin{eqnarray}
\mathcal{E}^{(\cN)}(u) &=& - \hbar c \Lambda^{2u} \frac{\cos\pi u}{2\pi} \iint \frac{d^2k_\perp}{(2\pi)^2} \int_0^\infty d\lambda \lambda^{1-2u}\ann
&\times& \frac{\partial}{ \partial \lambda} \ln \Psi_{\cN}(i\lambda c), \label{eq:EPlanar}
\end{eqnarray}
where $\Psi_{\cN}$ defines the electromagnetic energy spectrum, which can be found via the appropriate boundary condition.  Also, $k_\perp$ is the 2D wave vector and $\lambda=-i\omega/c$. The parameter $\Lambda$ with a wavenumber dimension is introduced to preserve the energy dimension of $\mathcal{E}(u)$.  To calculate the Casimir-Polder interaction, we take advantage of the idea developed by Lifshitz \cite{Lifshitz:1956:Ttmafbs} relying on media rarefication. Specifically, we take that the half space at $z<0$ to be described as $\varepsilon (\omega) = 1 + 4\pi L \alpha (\omega)$, where $L$ is the amount of atoms and $\alpha$ is the polarizability of single atom in this material. In the limit of  $L\to 0$ we obtain the energy $E^{(\cN)}$ per atom at a distance $a$:
\begin{equation}
E^{(\cN)} = -\lim_{L\to 0}\frac{1}{L}\frac{\partial \mathcal{E}^{(\cN)}}{\partial a},\label{eq:CP}
\end{equation}
where $\mathcal{E}^{(\cN)} = \mathcal{E}^{(\cN)}(L,u)$ is the zeta-regularized energy with regularization parameter $u$ for the configuration of $\cN$ planes and dielectric medium.

The zero-point energy $\mathcal{E}^{(\cN)}$ can be found from the standard boundary conditions when applied to Maxwell's electromagnetic equations to the system in Fig. \ref{fig:w1}. The energy is a summation of transverse electric (TE) and transverse magnetic (TM) contributions $E = E^{(\cN)}_\TE + E^{(\cN)}_\TM$, expressed as
\begin{eqnarray}
E^{(\cN)}_\TM &=& \int_0^\infty dy\int_0^1 dx  \alpha \left(\frac{xy}{d}\right)  \Gamma_\cN  \left(\eta \left(\frac{xy}{d}\right) \frac{1} x\right) (2-x^2),\ann
E^{(\cN)}_\TE  &=& \int_0^\infty dy\int_0^1 dx  \alpha \left(\frac{xy}{d}\right) \Gamma_\cN \left(\eta \left(\frac{xy}{d}\right) x\right) x^2. \label{eq:Main}
\end{eqnarray}
The TE and TM terms are conveniently given in terms of rescaled variables $y=2d\sqrt{k_\perp^2+\lambda^2}$ and $x=\lambda/\sqrt{k_\perp^2+\lambda^2}$. The two contributions have a common function 
\begin{gather*}
\Gamma_\cN \left(t\right) =  -\frac{y^3te^{-\frac{2a}dy}}{2\pi d^4} \left( 1 + t - e^{-y}f \frac{1 - f^{2(\cN-1)}}{1 - f^{2\cN}} \right)^{-1},\ann
f = \cosh  y + t \sinh y + \sqrt{\left(\cosh y +t \sinh y\right)^2 -1}.
\end{gather*}
Here $\eta = 2\pi\sigma/c$ is the dimensional conductivity. We note that the argument of $xy/d$ is a consequence of the frequency dependence in the response properties $\sigma(\omega)$ and $\alpha(\omega)$. Details in obtaining $E^{(\cN)}_\TM$, $E^{(\cN)}_\TE$, and $\Gamma_\cN \left(t\right)$ can be found in Appendix \ref{Sec:AppA}. 

The expressions for the TE and TM contributions can also be recast in a different form by further changing the variables $y = s\frac{d}{a}$  and  $z = \frac{xy}{d} = \frac{xs}{a}$. One finds 
\begin{eqnarray}
E^{(\cN)}_\TM &=& \int_0^\infty ds\int_0^{\frac{s}{a}} dz  \alpha (z) \widetilde{\Gamma}_\cN  \left(\frac{s\eta(z)}{za}\right) \left(2s^2-\left(za\right)^2\right),\ann
E^{(\cN)}_\TE  &=& \int_0^\infty ds\int_0^{\frac{s}{a}} dz  \alpha (z)   \widetilde{\Gamma}_\cN \left(\frac{za \eta(z)}{s}\right) \left(az\right)^2, \label{eq:Main1}
\end{eqnarray}
where
\begin{gather*}
 \widetilde{\Gamma}_\cN \left(t\right) =  -\frac{te^{-2s}}{2\pi a^3} \left( 1 + t - e^{-\frac{ds}{a}}\tilde{f} \frac{1 - \tilde{f}^{2(\cN-1)}}{1 - \tilde{f}^{2\cN}} \right)^{-1},\ann
\tilde{f} = \cosh  \frac{sd}{a}+ t \sinh \frac{sd}{a} + \sqrt{\left(\cosh \frac{sd}{a} +t \sinh \frac{sd}{a}\right)^2 -1}.
\end{gather*}
It turns out that some analytical results may be obtained in a more straight forward manner using either Eqs. (\ref{eq:Main}) or Eqs. (\ref{eq:Main1}) depending on the particular situation considered, as can be seen in what follows.  

The response properties of the atoms and layers are modeled as follows. The atomic polarizability is represented with an oscillator model 
\begin{equation}
\alpha (\lambda) = \frac{\alpha (0)}{1 + \frac{\lambda^2}{\lambda_a^2}}, \label{eq:manyosc}
\end{equation}
where $\alpha (0)$ is the static polaizability and $\lambda_a$ is the characteristic wavelength. Relevant parameters for several atoms, including H, $\mathrm{H}_2$, He, He* (excited He atom), Na, K, Rb, Cs and Fe, are given in Appendix \ref{Sec:AppC}. The graphene layer conductivity is represented via its universal value $\sigma_{gr}=e^2/4\hbar$. This is expected to be a good approximation when considering the Casimir-Polder interaction at large separations since  $\sigma_0$ is maintained over a relatively large frequency range $\hbar\omega\leq 3$ $eV$.  \cite{Falkovsky:2007:Sdogc,*Nair:2008:FSCDVToG}. The graphene conductivity is also described using a Drude-Lorentz (DL) model. Taking into account that $\sigma$ for a single graphene is very similar to the one for graphite in all frequency range except for small $\omega$, the graphene conductivity can be represented as a DL sum \cite{Djurisic:1999:Opog}, as can be seen in Appendix \ref{Sec:AppA}. 

We also note that throughout the paper we use $\hbar = c =1$, while in the final results the real units are restored.

\section{Asymptotic Relations}\label{Sec:AsymRel}
Several asymptotic expressions for the energy of the atom/multilayers system can be found when considering small or large atom/layer separations and/or constant graphene conductivity $\eta_0$. In the calculations, both forms in Eqs. (\ref{eq:Main}) and (\ref{eq:Main1}) are utilized depending on the particular case considered.

\subsection{$\cN$ planes, $a \to \infty$ }
The Casimir-Polder energy for an atom interacting with $\cN$ planes when the separation $a$ is large can be found using  Eq. (\ref{eq:Main}). Making the substitution $y = s\frac{d}{a}$ and then taking the limit of $a \to \infty$, it is found that $f\to 1$, while $\alpha \to \alpha (0)$, and $\eta \to  \eta (0)=\eta_0$. Performing the integration afterwards, one obtains the interaction energy 
\begin{equation}
E^{(\cN)}_{a\to\infty} = -\frac{3\alpha(0)}{8\pi a^4} q(\cN\eta_0). \label{eq:large}
\end{equation}
This result is very similar to the standard Casimir-Polder energy for an atom/perfect metal system (see, for example, \cite{Bordag:2009:ACE}), but with an additional factor $q(\cN\eta (0))$, which has the general expression
\begin{eqnarray}
q(x) &=& \frac{1}{6} + \frac{1}{2x^2} - \frac{1}{4x} + \frac{x^2}{2} - \frac{x}{4}\ann
& -& \frac{1}{2x^3}\ln (1+x) - \frac{x(x^2 - 2)}{2} \ln \left(1 + \frac{1}{x}\right). \label{eq:q}
\end{eqnarray}

\begin{figure}[ht]
	\centering
	\includegraphics[width=4.2truecm]{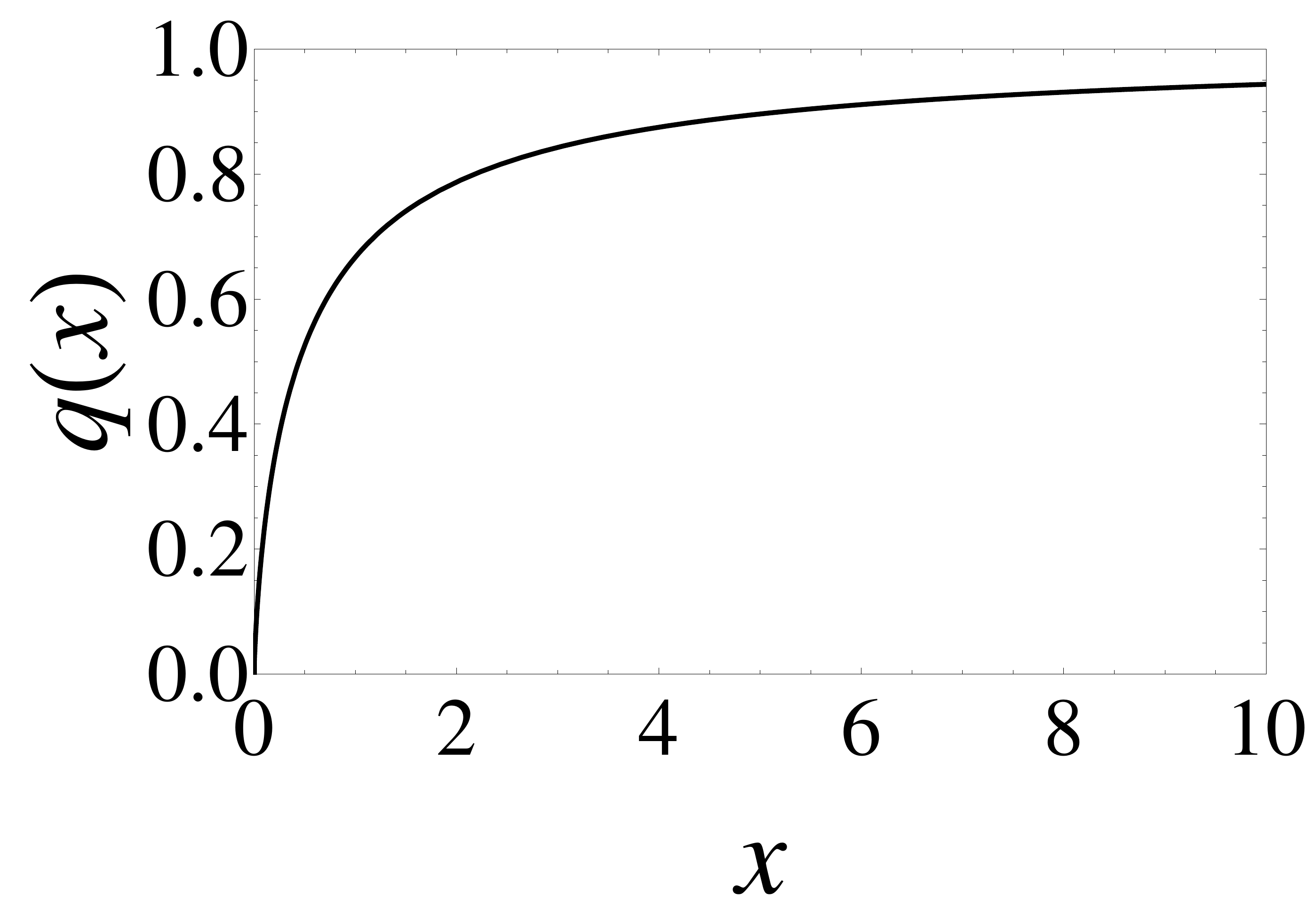}%
	\includegraphics[width=4.3truecm]{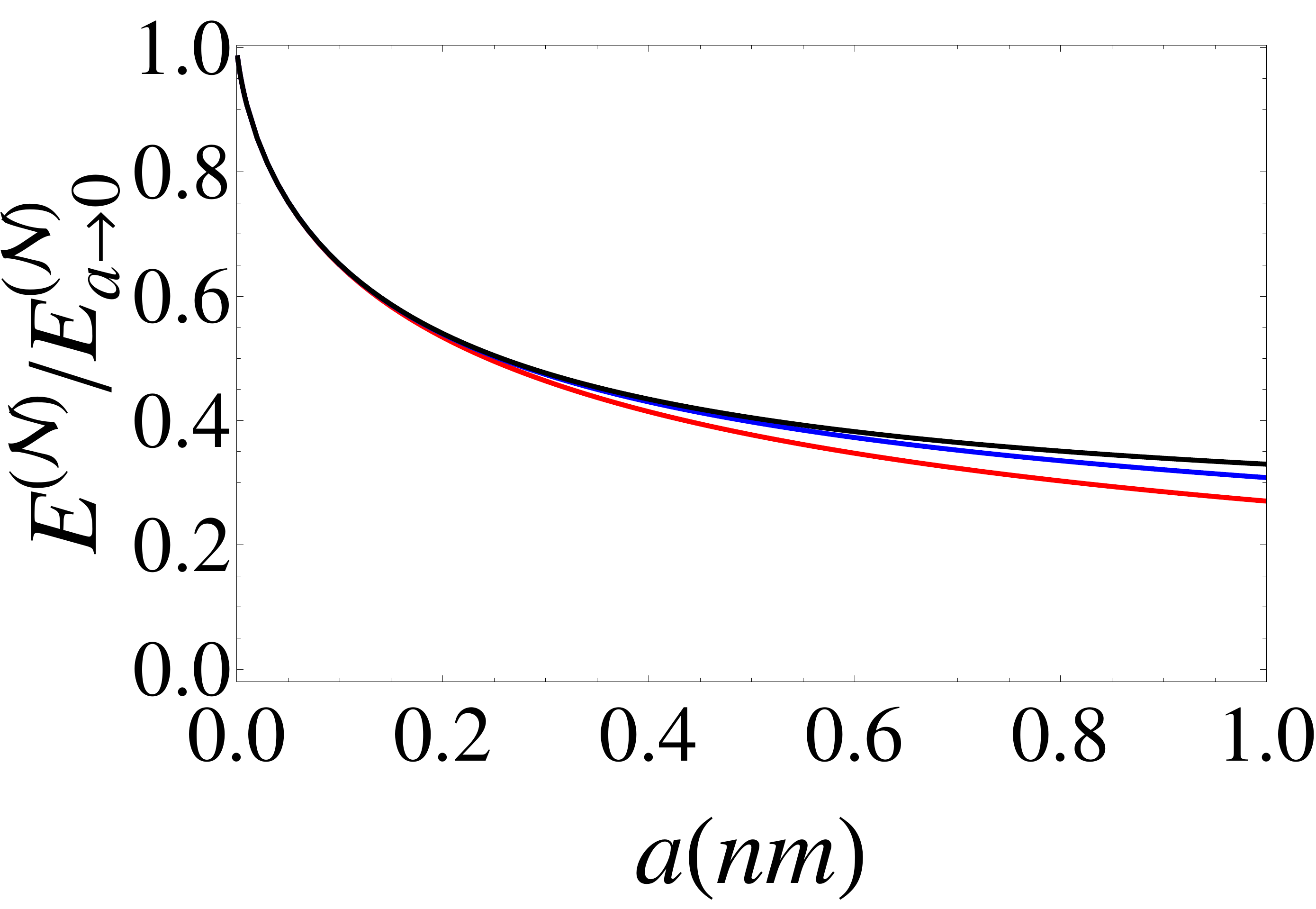}
	\caption{Left panel: the function $q(x)$ as a function of $x$. Right panel: the ratio  $E^{(\cN)}/E^{(\cN)}_{a\to 0}$ for Hydrogen atom and for different number of planes $\cN = 1,2,\infty$ (from button to top). We adopt the constant conductivity model with $\eta =\eta_{gr}$ and interplane distance $d=0.3354nm$.}\label{fig:w2}
\end{figure}

The explicit form of $q(x)$ enables us to examine other asymptotic behavior in terms of the magnitude of the constant conductivity and number of planes. For example, one finds $q(x)_{x \to \infty} = 1$. Therefore, for infinite number of planes, $\cN \to \infty$, or perfectly conducting planes, $\eta_0 \to \infty$, the standard Casimir-Polder energy is recovered.

The opposite limit can also be examined. Using Eq. (\ref{eq:q}), the small argument expansion gives
\begin{equation}
q(x)|_{x\to 0} = - \left(\frac{1}{8} + \ln x\right)x + O(x^2). 
\end{equation}
The above expression covers situations of small constant conductivity, which includes the case of graphene with estimated universal value $\eta_{gr}\approx 0.0114$. For example, the Casimir-Polder energy for an atom/single graphene is  $E^{(1)}_{a\to\infty, gr}=-\frac{3\alpha(0)}{8\pi a^4}0.05$. Thus the interaction is about 20 times smaller than the one involving a perfect metallic surface.

Eqs. (\ref{eq:large}) and (\ref{eq:q}) are also useful in better understanding how the energy of the atom/multilayers system can be manipulated. Increasing $\cN$, while $\eta_0$ is constant results in the same outcome when  $\eta_0$ is increased and 
$\cN$ is constant. Fig. \ref{fig:w2} (left panel) displays $q(x)$ as a function of $x$, which essentially traces the Casimir-Polder interaction energy when the atom is at a larger distance ($a \gg d$) from the $\cN$ stack of planes. 

\begin{figure}[htb]
	\centering
	\includegraphics[width=8truecm]{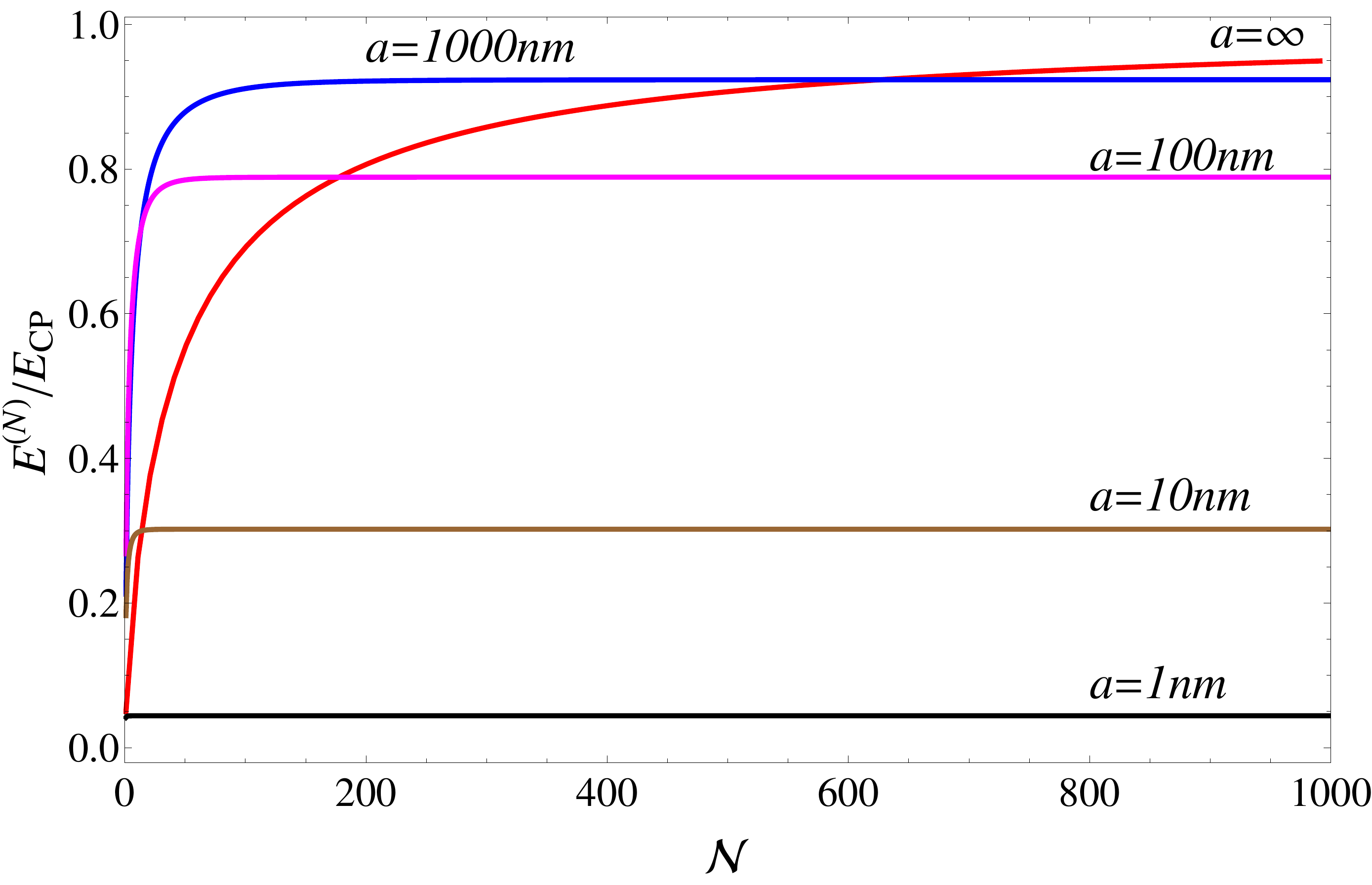}
	\caption{The Casimir-Polder energy of a Hydrogen atom, normalized by the Casimir-Polder energy atom/ideal metal, near a stack of graphene planes with conductivity described using DL model for $a = 10^n,\ n = 		0,1,2,3,\infty$. The interplane graphene separation is 
the equilibrium distance for graphite $d=0.3345nm$}\label{fig:w3}
\end{figure}

\subsection{$\cN$ planes, $a \to 0$ }
In the case of small atom/plane separations, the Casimir-Polder energy is mainly determined by the TM mode, while TE contribution is negligible, as can be seen from Eq. (\ref{eq:Main1}) after numerical calculations. Taking the limit of $ a \to 0$ in the TM contribution in Eq. (\ref{eq:Main1}) when $a\ll \eta (0)/\lambda_j, \eta (0)/\lambda_a$ ($\lambda_j$ are the DL characteristic frequencies as defined in Appendix \ref{Sec:AppB}), one finds
\begin{equation}
E_{a \to 0} = - \frac{1}{4\pi a^3} \int_0^\infty \alpha (z)dz. \label{eq:aSmall}
\end{equation}
This result was obtained in Ref. \cite{Zhou:1995:vdWarCioaeoaawmw}, where the authors have considered the vdW limit of  atom interacting with a perfect metal. It is interesting that the same result is obtained for the atom/multilayer configuration considered here. In addition to the energy being independent of $\cN$, there is no dependence on the magnitude of the conductivity providing the above discussed conditions, ($a\ll \eta (0)/\lambda_j, \eta (0)/\lambda_a$), are fulfilled.

Fig. \ref{fig:w2} (right panel) shows how $E^{\cN}/E_{a\to 0}$ changes as a function of $a$, where $E^{\cN}$ is calculated numerically using Eq. \ref{eq:Main}. It is evident that the interaction does not depend strongly on the number of planes involved. Also, Fig. \ref{fig:w2} (right panel) displays that $E^{\cN}/E_{a\to 0}$  does not change significantly as $a$ is increased showing that the $a \to \infty$ limit is quickly approached (after a few $nm$-s). 

\subsection{Large ($d\to \infty$) and small ($d \to 0$) interplane distances}

Asymptotic expressions can also be obtained by considering different limits of the interplane distance separation. Taking $d\to \infty$ while $a$ is finite in Eq. (\ref{eq:Main1}) the energy for a single plane ($\cN=1$) is recovered,  $E^{(1)}=E_\TM^{(1)}+E_\TE^{(1)}$. In addition, the limit of $d\to 0\ (d \ll a)$ in Eq. (\ref{eq:Main1}) is equivalent to the limit of $a\to \infty$ as we obtain that $E^{(\cN)}_{d\to 0} =  E^{(\cN)}_{a\to \infty}$. It is noted that these asymptotic limits hold for both models of the graphene conductivity providing that $d \gg \max (a,\lambda_j^{-1},\lambda_a^{-1})$ if $\sigma$ is described via the DL model and $d \gg \max (a,\lambda_a^{-1})$ if $\sigma$ is taken to be constant.

\begin{figure}[htb]
	\centering
	\includegraphics[width=8truecm]{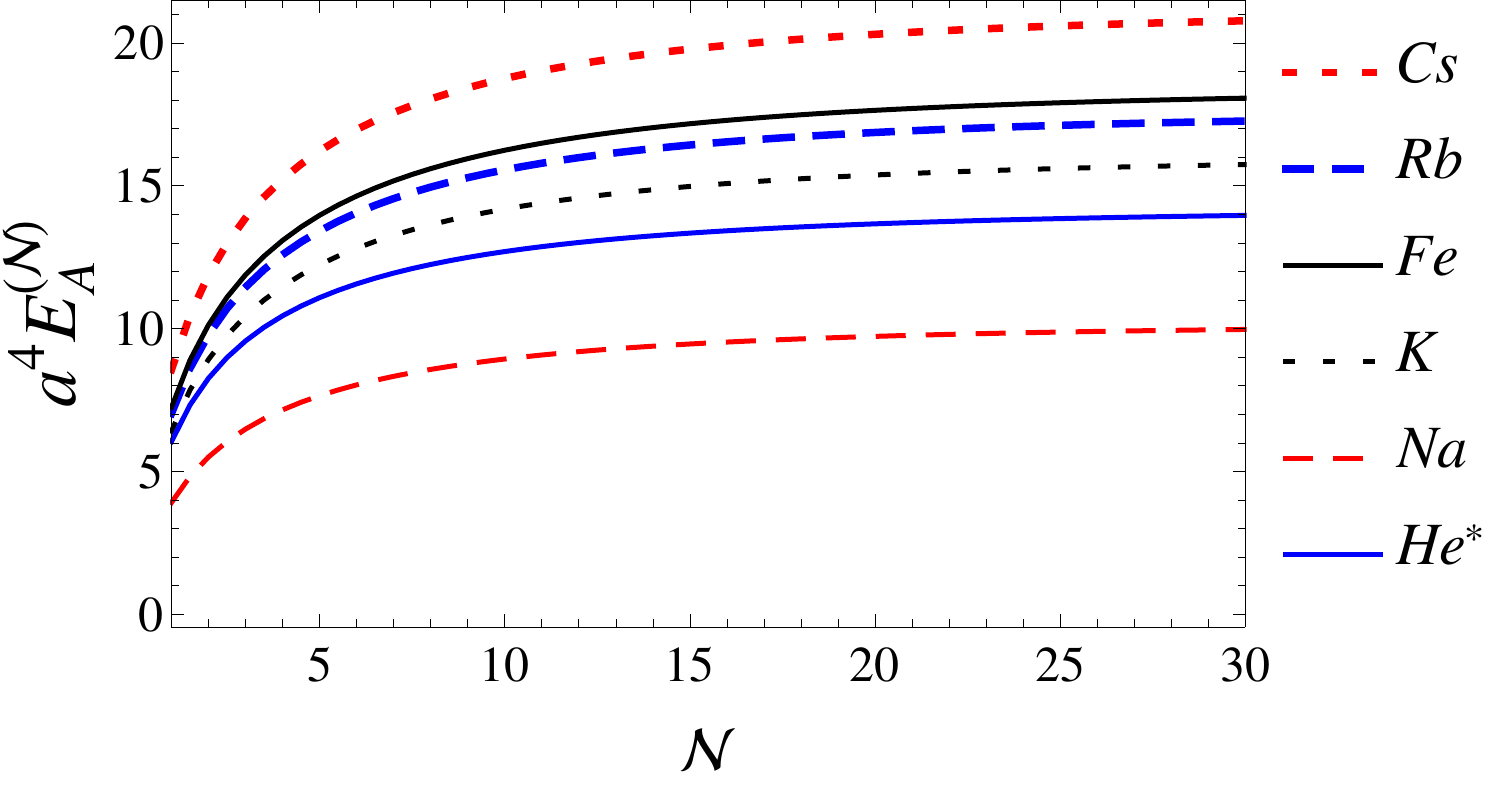}
	\caption{The Casimir-Polder energy multiplied by $a^4$ for several atoms as a function of number of graphene planes $\cN$. The conductivity for each graphene plane is taken into account using the DL model and the interplane distance is   $d=0.3345nm$ while $a = 100nm$.}\label{fig:w4n3}
\end{figure}

\section{Numerical simulations}\label{Sec:NumSim}

Analyzing the Casimir-Polder energy when the graphene sheets are described via the DL model (parameters are given in Appendix \ref{Sec:AppB}) requires numerical calculations. Such calculations are also needed when evaluating the interaction beyond the asymptotic limits for $\eta=\eta_0$ discussed previously. Taking the parameters for several atoms (Appendix \ref{Sec:AppC}) $E^{(\cN)}$ can be calculated as a function of the $a$ and $d$ separations and number of planes $\cN$ using Eq. (\ref{eq:Main}).

In Fig. \ref{fig:w3}, the Casimir-Polder energy for a Hydrogen atom is shown as a function of the number of graphene planes $\cN$ for different distances $a = 10^n$ $nm ,\ n = 0,1,2,3,\infty$ where the graphene conductivity is taken using the DL model and $\eta_{gr}$. Here the equilibrium for graphite interplane distance is used such that  $d=0.3345nm$ \cite{Girifalco:1956:EoCCatPEFotGS}.  

\begin{figure}[htb]
	\centering
	\includegraphics[width=8truecm]{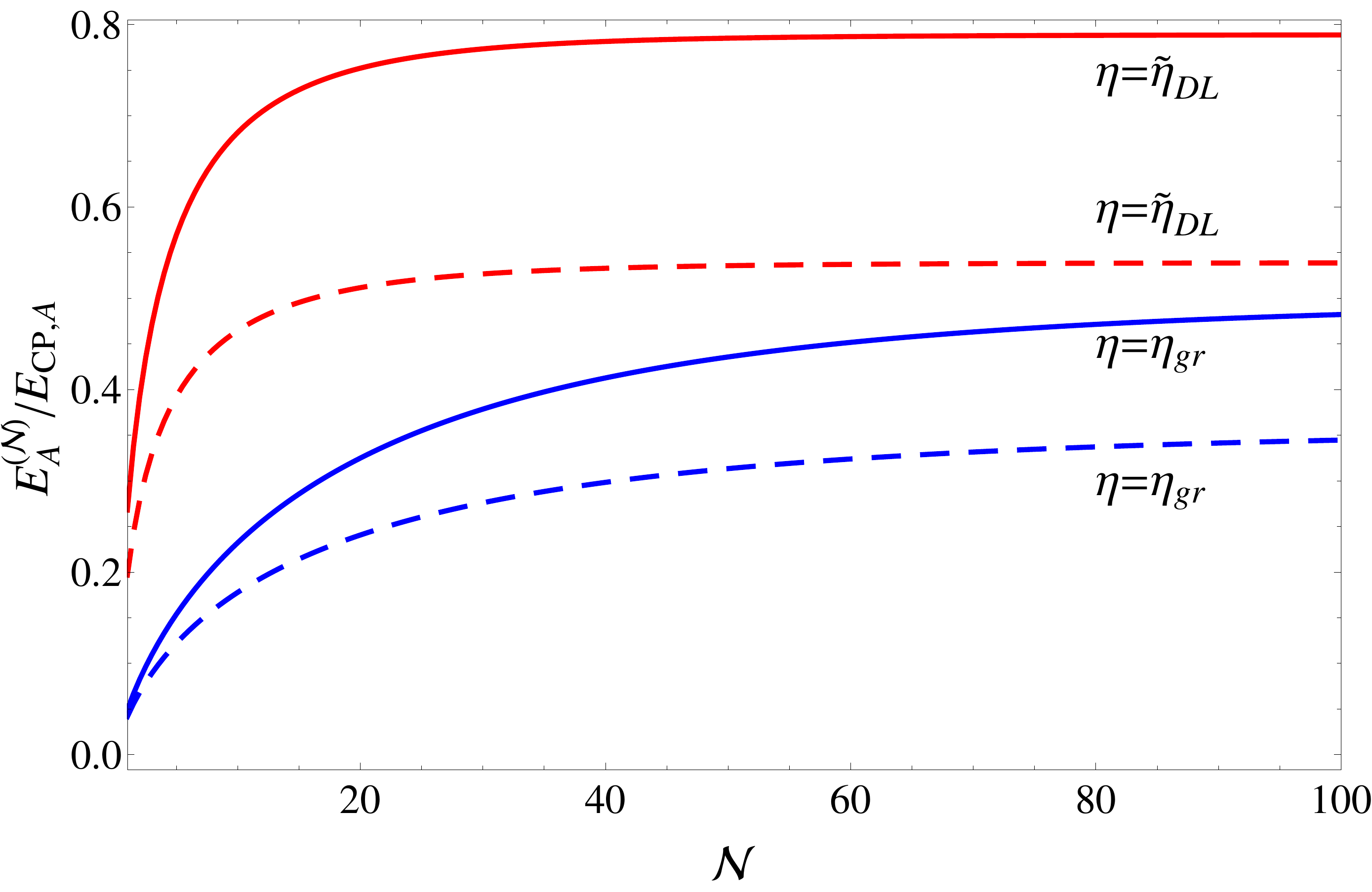}
	\caption{The Casimir-Polder energy for Hydrogen (solid lines) and Rb (dashed lines) atom near a stack of graphene planes with DL (upper two curves) and $\eta_{gr}$ (lower two curves) conductivities. Here $d=0.3345nm$ and $a = 100nm$.}\label{fig:w5}
\end{figure}

Fig. \ref{fig:w3} shows that for all $a$ the normalized $E^{(\cN)}/E_{CP}$ increases up to a certain $\cN$, after which there is no change. Depending on $a$, reaching the constant  $E^{(\cN)}/E_{CP}$ happens at different $\cN$. Specifically, for smaller atom/plane separations, the constant $E^{(\cN)}/E_{CP}$ is achieved at much smaller values of $\cN$ as opposed to larger $a$, where the constant behavior is achieved when there are many planes in the stack. 

We also consider how the different atoms affect the interaction in terms of their characteristics, specified in Appendix \ref{Sec:AppC}. Fig. \ref{fig:w4n3} shows that all atoms affect the energy in a similar way as a function of $\cN$. The rescaled $E^{\cN}$ increases as a function of $\cN$ until a plateau is reached, which corresponds to the asymptotic limit of large $\cN$ in Eq. (\ref{eq:large}), where $q(\cN \eta_0) \to 1$ due to the large argument. The magnitude of the plateau region is essentially determined by $\alpha(0)$. Therefore, we see that atoms with larger $\alpha(0)$, such as Cs, have bigger energy, as opposed to the ones with smaller $\alpha(0)$, such as Na. 

Is is also interesting to compare how the different models for the graphene conductivity affect the Casimir-Polder interaction. Fig. \ref{fig:w5} shows the energy for a Hydrogen atom at a distance $a=100nm$ when each graphene plane is described using both models. Although the characteristic behavior as a function of $\cN$ is the same, the $E^{\cN}/E_{CP}$ is smaller when $\eta=\eta_{gr}$. In addition, the constant region is achieved faster for the DL model as compared to constant conductivity.

The Casimir-Polder interaction for the considered system here can further be analyzed by separating the retarded and non-retarded regimes. This can be achieved by casting the total energy from Eqs. (\ref{eq:Main}) via $y=s \frac{d}{a}$ in the form
\begin{equation}
E^{(\cN)} = - \frac{C_3(a,\cN)}{a^{3}}, \label{eq:C3}
\end{equation}
where  $C_3(a,\cN)$ is the vdW coefficient, which depends on the number of planes and the atom/plane separation $a$. The $1/a^3$ dependence and $C_3$ constant are characteristic for the non-retarded vdW regime, typically for $a<5$ $nm$. For larger $a$ however, retardation effects become important and the $1/a^3$ behavior is no longer valid.

\begin{figure}[ht]
	\includegraphics[width=8.5truecm]{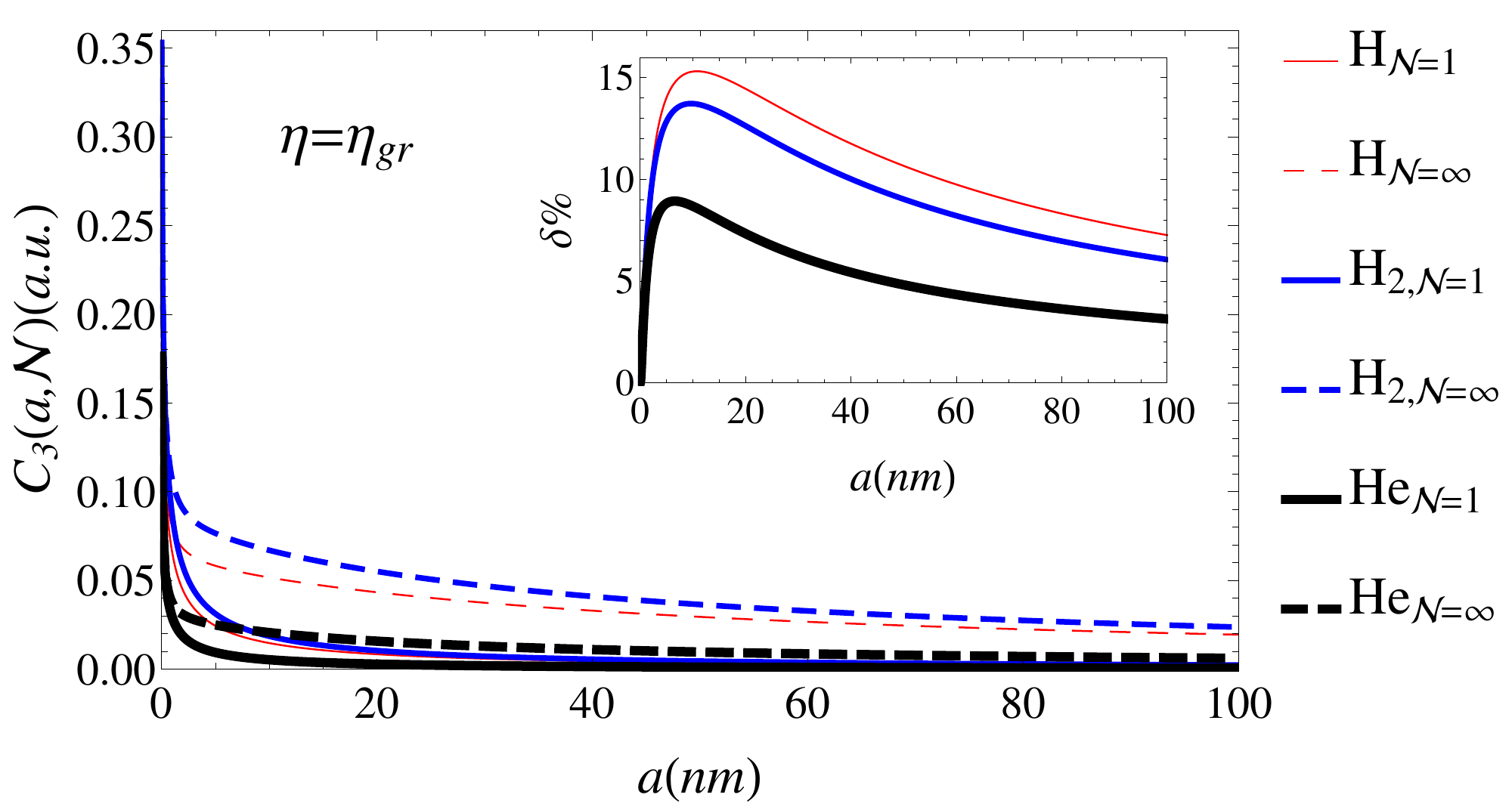}
	\includegraphics[width=8.5truecm]{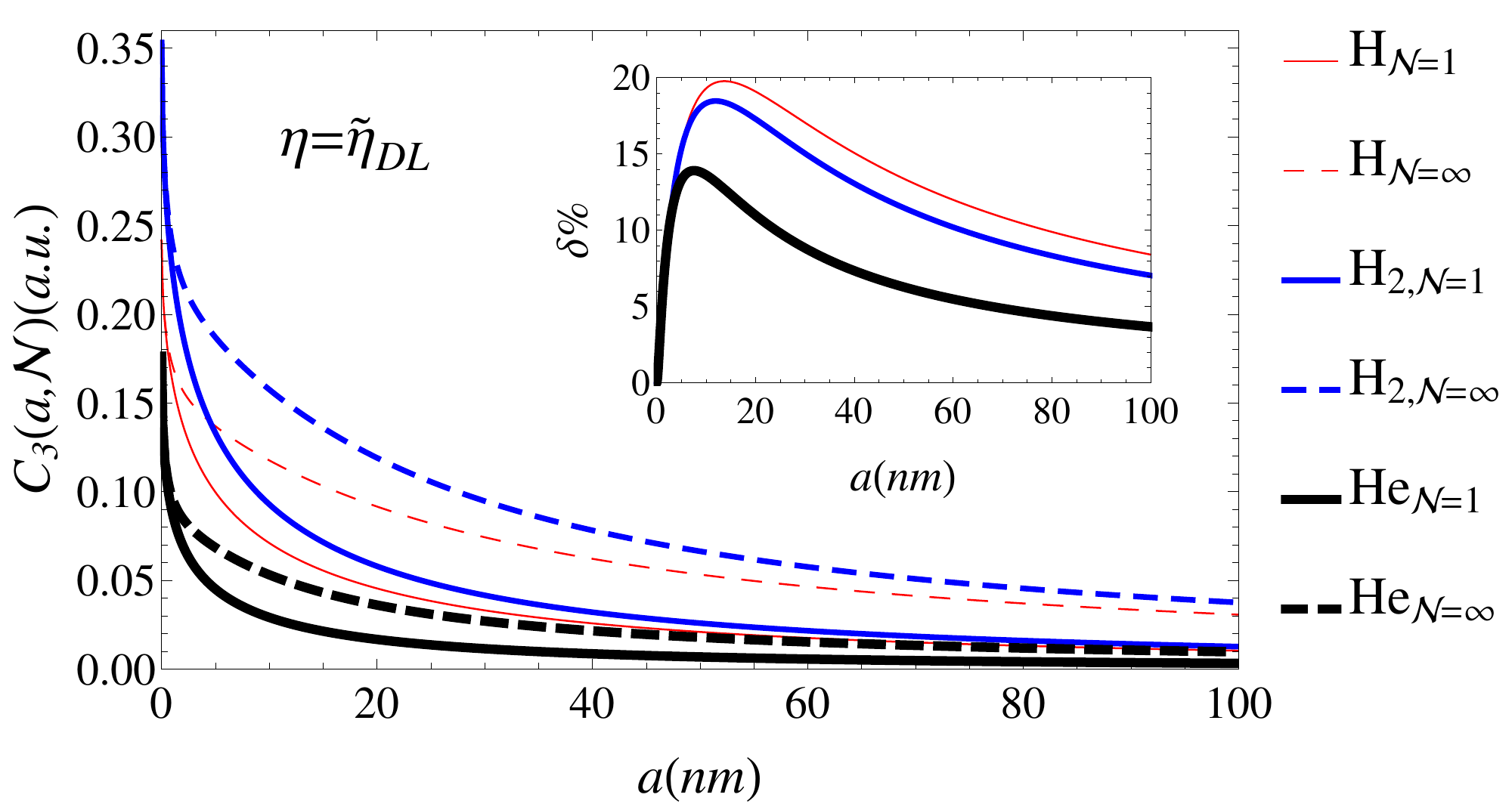}
	\caption{The vdW coefficient $C_3(a,\cN)$ for several light atoms and  for two models of the conductivity. Solid lines: $\cN =1$ and dashed lines: $\cN =\infty$. The insert is the relative difference given by Eq. (\ref{eq:delta}) as a function of $a$ separation. }\label{fig:DLCC}
\end{figure}

\begin{figure}
	\includegraphics[width=8.5truecm]{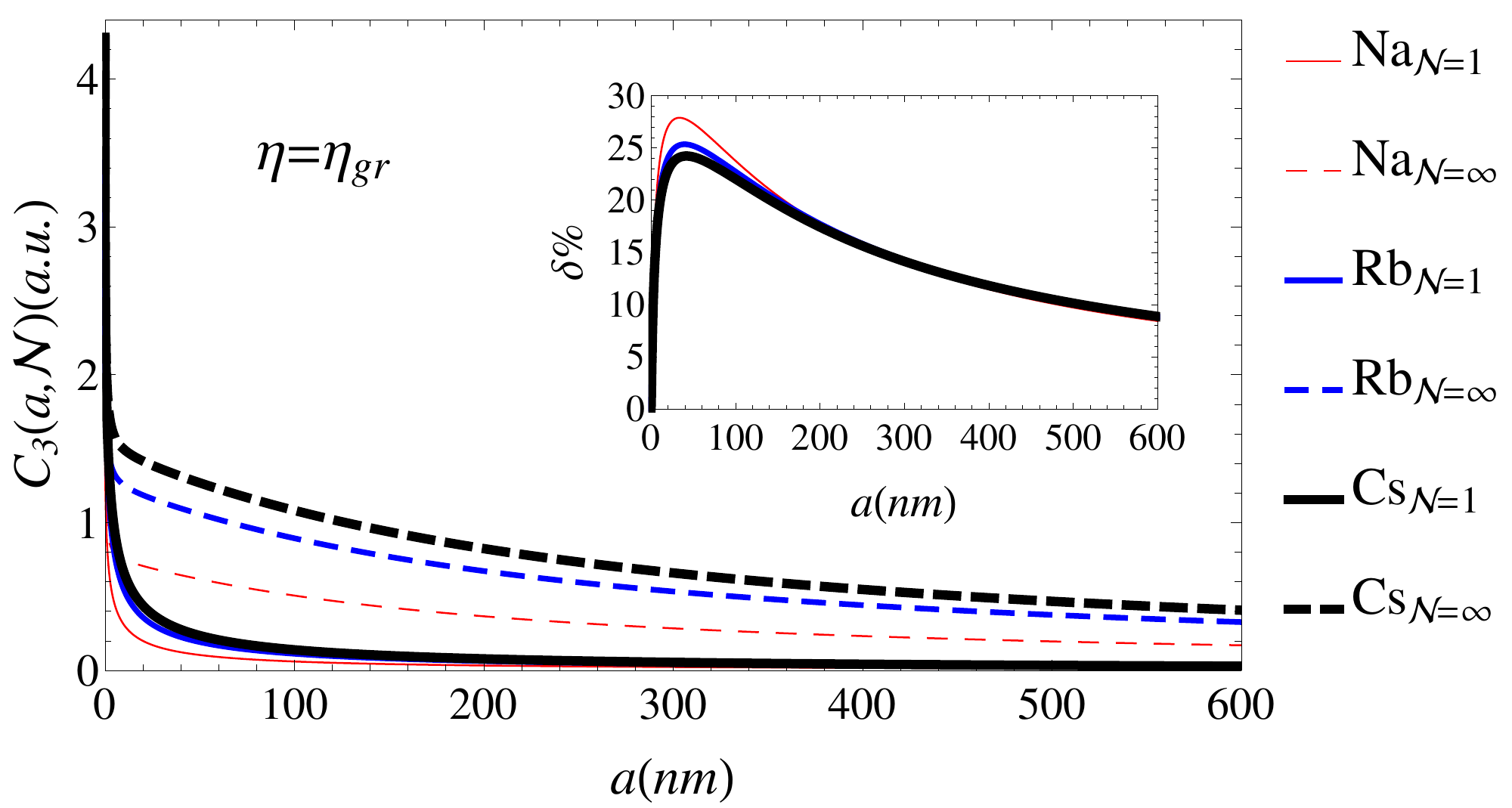}
	\includegraphics[width=8.5truecm]{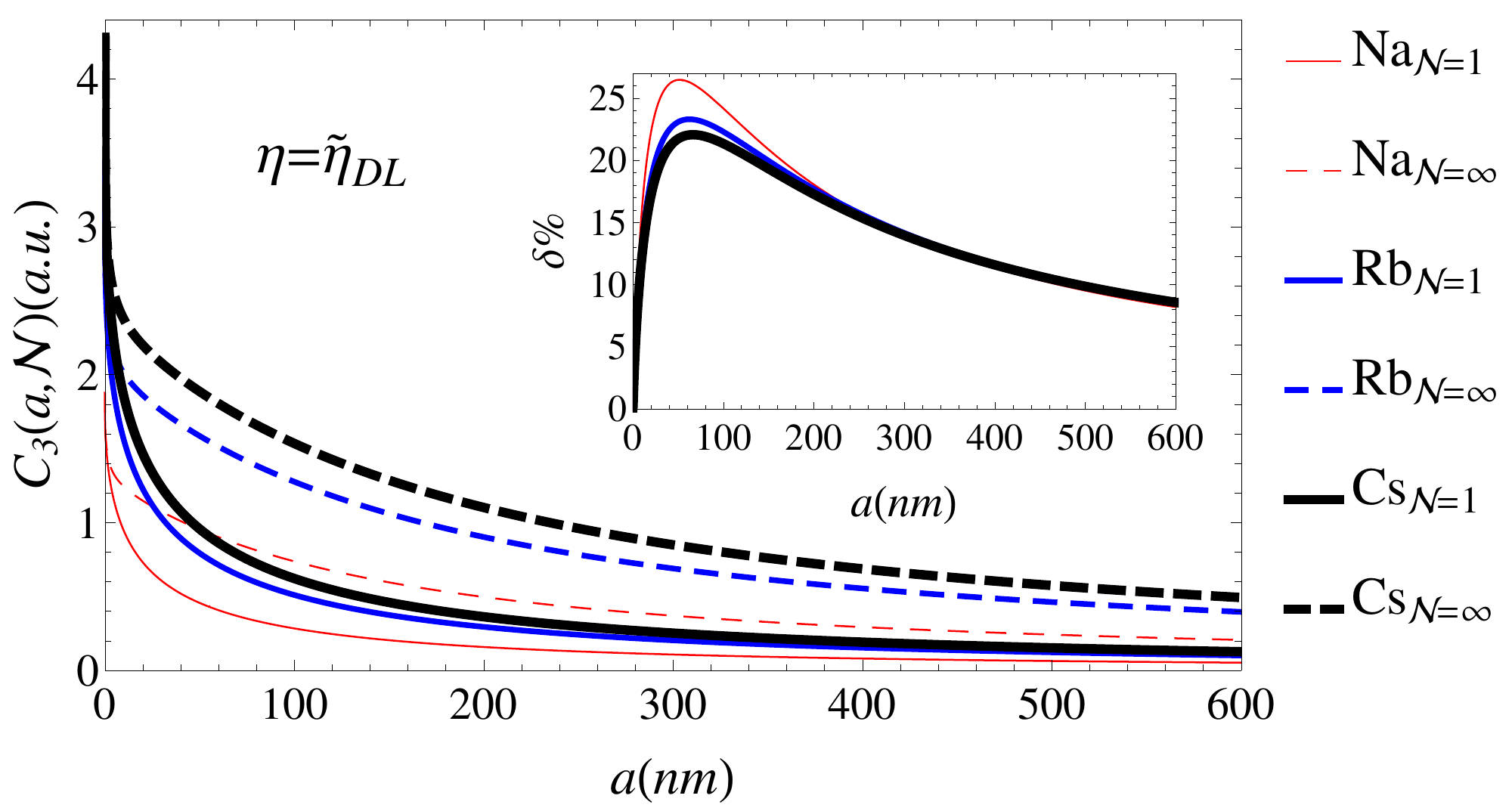}
	\caption{The vdW coefficient $C_3(a,\cN)$ for several heavy atoms and  for two models of the conductivity. Solid lines: $\cN =1$ and dashed lines: $\cN =\infty$. The insert is relative the difference given by Eq. (\ref{eq:delta}) as a function of separation $a$. }\label{fig:DLCC2}
\end{figure}

Investigating $C_3(a,\cN)$ dependence on $a$ is instructive for the understanding of the importance of the relativistic effects, while the $\cN$ dependence is indicative of how the size of the planar stack influences the interaction. In Figs. \ref{fig:DLCC} and \ref{fig:DLCC2}, we show the vdW coefficient for several atoms as a function of $a$ for the two models of the graphene conductivity. One notes that in both cases, $C_3(a,\cN)$ decreases as $a$ grows, however, this decay is stronger for $\eta=\eta_{gr}$, which is the reason for the smaller vdW coefficient when compared with the one found with $\eta=\eta_{DL}$. The strength of the atomic polarizability and number of planes determine the overall value of the vdW coefficient. For larger $\alpha_0$ (Na, K, Rb, CS), $C_3$ has bigger magnitude when compared to the $C_3$  for H, H$_2$, He. Similarly, larger $\cN$ results in larger vdW coefficients.

Examining the large $a$ behavior in Figs. \ref{fig:DLCC} and \ref{fig:DLCC2} shows that $C_3(a,\cN)\sim 1/a$. Therefore, the overall behavior of the interaction energy is $E\sim 1/a^4$, which coincides with the asymptotic limit discussed in Sec. \ref{Sec:AsymRel}. Figs. \ref{fig:DLCC} and \ref{fig:DLCC2}  show that the $1/a$ behavior becomes apparent for shorter separations when the number of planes is small and for the constant graphene conductivity model.

The comparison of the vdW coefficients for the different atoms calculated with the two models for the graphene conductivity can further be examined by calculating the following difference
 \begin{equation}
\delta (a)= \frac{C_3(a,\infty) - C_3(a,1)}{C_3(0,1)}100\% . \label{eq:delta}
\end{equation} 
The above expression gives means to obtain how the difference between the vdW coefficients for $\cN=1$ and $\cN=\infty$ normalized to the coefficient at $a=0$ for one graphene plane changes as a function of $a$. The insert in Figs. \ref{fig:DLCC} and \ref{fig:DLCC2}  shows that $\delta(a)$ has non-monotonic behavior. It experiences a maximum point at a certain separation $a$. For atoms with smaller polarizability $\delta_{max}$ is found at shorter separations. For atoms with larger polarizability $\delta_{max}$ is fairly insensitive to the $a$ separation.

\section{Conclusion}\label{Sec:Conclusion}

The Casimir-Polder interaction is a ubiquitous force in any atom/substrate system. Due to recent experimental advances in materials at the nanoscale, atom/multilayered systems composed of graphene sheets are of particular interest. In this work we present a perspective of Casimir-Polder effects using summation of zero-point energies via the zeta-function regularization technique. The zero-point energies are found by solving the boundary conditions of the electromagnetic spectrum which allows the utilization of different models for the response properties of the atoms and layers.

The asymptotic derivation for large separations between the atom and the graphene stack shows that the Casimir-Polder energy takes a form similar to the one for atom/single graphene and a numerical factor $q(\cN\sigma (0))$. Thus the interaction can be increased, for example, either by increasing $\sigma(0)$ or the number of graphenes in the stack $\cN$. The small separation limit between the atom and the graphenes recovers the expression for the atom/single graphene configuration. The large and small inter-graphene distances are also considered. The derived asymptotic relations compare well with the numerical calculations of the derived Casimir-Polder expression. The numerical calculations also show that the constant conductivity description results in a smaller magnitude interaction as opposed to the Drude-Lorentz model for $\sigma(\omega)$.

We also consider the non-retarded regime by calculating the vdW coefficient $C_3(a,\cN)$, which depends on the atomic separation and the number of planes. Again, it is found that $C_3(a,\cN)$ is strongly dependent on the model for $\sigma(\omega)$ as $\sigma_{gr}$ results in a smaller coefficient when compared to the calculations with the DL model. The atomic polarizability also affects the interaction showing that atoms with larger $\alpha(0)$ have stronger Casimir-Polder interaction.

This work shows that the applicability of the zeta-regularization technique can be expanded to atom/multilayer systems. The asymptotic expressions help us gain insight into the basic science of the Casimir-Polder interaction. The calculations for the various atoms and models for the graphene conductivity are also useful to find ways how to manipulate this interaction.

\begin{acknowledgements}
N.K. and R.K. were supported in part by the Russian Foundation for Basic Research Grant No. 16-02-00415-a. L.M.W. acknowledges financial support from the US Department of Energy under Grant No. DE-FG02-06ER46297.
\end{acknowledgements}

\appendix
\section{The energy of an atom near the stack of $\cN$ conductive planes}\label{Sec:AppA} 

Due to the planar symmetry of the system in Fig. \ref{fig:w1}, the electric ($\mathbf{E}$) and magnetic ($\mathbf{H}$) fields are represented by the following
\begin{equation}
	\mathbf{E} = \mathbf{e}(z) e^{i k_x x + i k_y y - i\omega t },  \ \mathbf{H} = \mathbf{h}(z) e^{i k_x x + i k_y y - i\omega t}.
\end{equation}
Taking into account that Ohm's law is satisfied on each conductive surface, the boundary conditions can be written in a decoupled form for the TE and TM contributions. 

\subsubsection{\TM\ mode, $H_z =0$}

The TM electromagnetic modes are found by solving the Maxwell equations with $\mathbf{E}$ and $\mathbf{H}$  for the system in Fig. \ref{fig:w1}. The dielectric medium at $z<0$ is considered to have frequency dependent response properties $\varepsilon = \varepsilon (\omega),\mu = \mu (\omega)$ and each 2D layer has the same conductivity $\sigma(\omega)$, while the domains between $z=0$ and $z=a+(\cN -1)d$ are with $\varepsilon =1,\mu =1$. The respective boundary conditions are 
\begin{eqnarray}
	\left[e_z'\right]_{z = a+jd} &=& 0,\ann
	\left[e_z\right]_{z = a+jd} &=& -\frac{4\pi i \sigma }{\omega} e_z' = - \frac{2\pi\sigma}{c\kappa} e_z', \ann
	\left[e_z'\right]_{z = 0} = 0 &,&\ \left[\varepsilon e_z\right]_{z = 0} = 0,
\end{eqnarray}
 where $[f]_{z} = f(z-0)-f(z+0)$. 
  
Thus there are  $2\cN+2$ coupled equations, whose main determinant can be written on the imaginary axis $\omega = i \lambda$ according to Eq. (\ref{eq:EPlanar}),
\begin{equation}
\triangle_\TM =
\begin{vmatrix}
Z_1 & B_0 & 0&\dots& 0&0 \\
0 & A_0& B_1 &\dots& 0  &0\\
0 & 0 & A_1 &\dots& 0 &0\\
\vdots&\vdots&\vdots&\ddots&\vdots&\vdots\\
0 & 0 & 0 &\dots& A_{\cN-2} & B_{\cN-1} \\
Z_2 & 0 & 0 &\dots& 0 & A_{\cN-1}
\end{vmatrix},\label{eq:Det}
\end{equation}
This $(\cN+1)\times (\cN+1)$ determinant is in a block-diagonal form with elements
\begin{gather}
B_l=
\begin{pmatrix} 
e^{(l-1) p} & -e^{-(l-1) p} \\
e^{(l-1) p} & e^{-(l-1) p}
\end{pmatrix},
B_0=
\begin{pmatrix} 
e^{-\oa p} & -e^{\oa p} \\
e^{-\oa p} & e^{\oa p}
\end{pmatrix},\ann
A_j =
\begin{pmatrix}
-e^{j p} & e^{-j p} \\
(-\frac{2\pi\sigma}{c}-1) e^{j p} & (\frac{2\pi\sigma}{c}-1) e^{-j p}
\end{pmatrix},\ann
Z_1=
\begin{pmatrix}
-\frac{\kappa_\varepsilon}{\kappa} & 0 \\
-\varepsilon & 0 
\end{pmatrix},
Z_2=
\begin{pmatrix}
0 & -e^{-(\cN-1) p} \\
0 & e^{-(\cN-1)p}
\end{pmatrix},
\end{gather}
where $l=(1,\cN-1)$, $j=(0, \cN -1)$. Also, $p=d\kappa,\ \oa = a/d$, $\kappa_\varepsilon= \sqrt{k_\perp^2 + \varepsilon \mu \lambda^2}$, and $\kappa= \sqrt{k_\perp^2 + \lambda^2}$.

The diagonalization of the matrix yields 

\begin{equation}
\triangle_\TM = \det\left[ Z_1 + (-1)^{\cN}C(\oa p)C^{(\cN-1)}(p)Z_2\right],
\end{equation}
where 
\begin{equation*}
C(x) = B_lA^{-1}_l = 
\begin{pmatrix}
-\cosh x  - \frac{2\pi\sigma}{c} \sinh x & \sinh x \\
\frac{2\pi\sigma}{c} \cosh x+\sinh x & -\cosh x
\end{pmatrix}.
\end{equation*}
To calculate the  ($\cN-1$) power of the matrix $C$, we represent it in Jordan form $C=T J T^{-1}$, where 
\begin{equation}
T = \begin{pmatrix}
\frac{-\sinh p}{\cosh p - f^{-1}} & \frac{-\sinh p}{\cosh p -f}\\
1&1
\end{pmatrix},
J= 
-\begin{pmatrix}
f&0\\
0&f^{-1}
\end{pmatrix},
\end{equation}
and $f = \sqrt{(\cosh p + b_1 \sinh p)^2 -1} +(\cosh  p+\frac{\eta \kappa}{\lambda}  \sinh p)$.  

For the calculation of the Casimir-Polder force one needs the renormalized spectrum, which requires finding $\triangle_\TM^\infty$ at the limit of $a\to\infty$ and $d\to \infty (p\to\infty)$. Thus the renormalized determinant reads 
\begin{gather}
\frac{\triangle_\TM}{\triangle_\TM^\infty} = -\frac{e^{-p(\cN-1)}}{f^{\cN-2}(1+\frac{\eta \kappa}{\lambda})^\cN} \left\{ e^{-p} \frac{1 - f^{2(\cN-1)}}{1 - f^2}\right.\ann
\left. - \frac{1 - f^{2\cN}}{f(1 - f^2)} \left(1+\frac{\eta \kappa}{\lambda} + \frac{\eta \kappa}{\lambda} e^{-2\oa p} \frac{\kappa_\varepsilon - \varepsilon\kappa}{\kappa_\varepsilon + \varepsilon\kappa}\right)\right\}.\label{eq:phi}
\end{gather}
One notes that for $\varepsilon = \mu = 1$, we obtain the result for $\cN$ planes, as found in \cite{Khusnutdinov:2015:Cefasocp}. 

Finally, after changing variables in  Eq. (\ref{eq:EPlanar}) to spherical coordinates $k_x = \kappa \sin\theta \cos\varphi, k_y = \kappa \sin\theta \sin\varphi, \lambda = \kappa \cos\theta\ (\kappa = \sqrt{k_x^2 + k_y^2 + \lambda^2})$, $y = 2p = 2d\kappa, \lambda = \kappa x, x=\cos\theta$, the Casimir-Polder energy can be written as 
\begin{equation}
\mathcal{E}^{(\cN)}_\TM =  \frac{Q^{(\cN)}_\TM}{d^3},
\end{equation}
where  
\begin{gather}
Q^{(\cN)}_\TM =  \int_0^\infty y^2dy\int_0^1 dx \ln \Phi^{(\cN)}_\TM\left(\frac{\eta}{x}\right), \ann
\Phi^{(\cN)}_\TM(t) =  \frac{-e^{-p(\cN-1)}}{32\pi^2 f^{\cN-2}(1+t)^\cN} \left\{ e^{-p} \frac{1 - f^{2(\cN-1)}}{1 - f^2}\right.\ann
\left. - \frac{1 - f^{2\cN}}{f(1 - f^2)} \left(1+t + t e^{-2\oa p} \frac{\kappa_\varepsilon - \varepsilon\kappa}{\kappa_\varepsilon + \varepsilon\kappa}\right)\right\},\ann
f = \sqrt{(\cosh p + t \sinh p)^2 -1} +\cosh  p+t  \sinh p.\label{eq:QTM}
\end{gather}

Rarefying the media $\varepsilon(\omega)=1 + 4\pi L \alpha (\omega)$ and $\mu (\omega) = 1 + 4\pi L \beta (\omega)$ ($\alpha(\omega)$ -- atomic polarizability; $\beta (\omega)$ -- atomic dynamic magnetic susceptibility) leads to the Casimir-Polder energy. For an atom with trivial magnetic properties such as $\mu=1$, the media rarefication is only for $\varepsilon(\omega)$. Below we provide both situations
\begin{equation}
E^{(\cN)}_\TM = \int_0^\infty dy\int_0^1 dx  \alpha \left(\frac{xy}{2d}\right) (2-x^2) \Gamma_\cN  \left(\frac \eta x\right),
\end{equation}
\begin{equation}
\hat E^{(\cN)}_\TM = - \int_0^\infty dy\int_0^1 dx  \beta \left(\frac{xy}{2d}\right)x^2\Gamma_\cN \left(\frac \eta x\right),
\end{equation}
where  
\begin{equation*}
\Gamma_\cN \left(t\right) =  -\frac{y^3te^{-\frac{a}dy}}{32\pi d^4} \left( 1 + t - e^{-\frac{y}{2}}f \frac{1 - f^{2(\cN-1)}}{1 - f^{2\cN}} \right)^{-1}.
\end{equation*}

\subsubsection{\TE\ mode, $E_z =0$}

Obtaining the electromagnetic spectrum for the TE modes follows a similar procedure for the appropriate boundary conditions
\begin{eqnarray}
\left[h_z\right]_{z = jd} &=& 0,\ann
\left[h_z'\right]_{z = jd} &=& 4\pi i \sigma \omega h_z = - \frac{2\eta \lambda }{\kappa}\kappa h_z, \ann
\left[h_z'\right]_{z = 0} &=& 0,\ \left[\mu h_z\right]_{z = 0} = 0. 
\end{eqnarray}
The determinant of the coupled equations has the same form as in Eq. (\ref{eq:Det}) and 
\begin{equation}
B_0=
\begin{pmatrix} 
e^{-\oa p} & e^{\oa p} \\
e^{-\oa p} & -e^{\oa p}
\end{pmatrix},
Z_1=
\begin{pmatrix}
-\frac{\kappa_\varepsilon}{\kappa} & 0 \\
-\mu & 0 
\end{pmatrix}.
\end{equation}
Executing the variable change to spherical coordinates, the energy is found as  
\begin{equation}
\mathcal{E}^{(\cN)}_\TE =  \frac{Q^{(\cN)}_\TE}{d^3},
\end{equation}
where  
\begin{equation}
Q^{(\cN)}_\TE =   \int_0^\infty y^2dy\int_0^1 dx \ln \Phi^{(\cN)}_\TE\left(\eta x\right),
\end{equation} 
and 
\begin{gather}
\Phi^{(\cN)}_\TE(t) = -\frac{e^{-p(\cN-1)}}{32\pi^2f^{\cN-2}(1+t)^\cN} \left\{ e^{-p} \frac{1 - f^{2(\cN-1)}}{1 - f^2} \right.\ann
\left. - \frac{1 - f^{2\cN}}{f(1 - f^2)} \left(1+t + t e^{-2\oa p} \frac{\mu \kappa - \kappa_\varepsilon}{\mu \kappa + \kappa_\varepsilon}\right)\right\}.
\end{gather}
Rarefying the dielectric medium yields 
\begin{eqnarray}
\hat E^{(\cN)}_\TE &=& - \int_0^\infty dy\int_0^1  dx  \beta \left(\frac{xy}{2d}\right)(2-x^2) \Gamma_\cN \left(\eta x\right),\ann
E^{(\cN)}_\TE &=& \int_0^\infty dy\int_0^1 dx  \alpha \left(\frac{xy}{2d}\right)x^2\Gamma_\cN \left(\eta x\right),
\end{eqnarray}
where we marked by hat the magnetic contributions. 

We finally note that there is a simple relation between the $\Gamma_{\cN}$ function and the reflection coefficients $r_{TM,TE}$ of the atom/multilayered system as follows, 
 \begin{equation}
 \Gamma_\cN  \left(\frac \eta  x,y\right) = r_\TM, \ \Gamma_\cN \left(\eta x,y\right) = - r_\TE.
 \end{equation}
Thus the total Casimir-Polder energy can be expressed as
 \begin{eqnarray}
 E^{(\cN)} &=& \int_0^\infty dy\int_0^1 dx \left\{2\left[\alpha r_\TM + \beta r_\TE\right]\right. \ann
&-&\left. x^2 \left[\alpha + \beta\right] \left[r_\TM + r_\TE\right]\right\},
 \end{eqnarray}
which has the same structure as the energy obtained via the Lifshitz approach for an atom/substrate system (Eq. (16.91) in 
 \cite{Bordag:2009:ACE}.

\section{Dielectric Response Properties}

\subsection{The Drude-Lorentz model of conductivity}\label{Sec:AppB}

The optical conductivity of graphene sheet is very close to the one for an isolated graphene over a wide range of frequencies \cite{Marinopoulos:2004:isotoaawdrog}. The results for graphene have been mapped to a Drude-Lorentz model consisting of  a Drude term and seven Lorentz oscillators according to \cite{Djurisic:1999:Opog} 
\begin{equation}
\sigma_{DL} (\omega) = \frac{f_0 \omega_p^2}{\gamma_0 - i\omega} + \sum_{j =1}^7\frac{i \omega f_j \omega_p^2}{ \omega^2 - \omega_j^2 + i\omega \gamma_j }.\label{eq:DL}
\end{equation}

The graphene DL conductivity is obtained from  Eq.  (\ref{eq:DL}) for the 3D graphite by multiplying it with $2\pi a/c$ ($a$ is a characteristic distance typically taken as the interplanar distance of graphite). The expression is given on the  imaginary axis $\omega = i \lambda c \ (k=i\lambda)$ as follows:
\begin{equation}
\eta_{DL}(\lambda)  = \frac{\eta_0 \tilde\gamma_0}{\tilde\gamma_0 + \lambda} + \sum_{k = 1}^7\frac{\lambda \eta_k \tilde\gamma_k }{ \lambda^2 + \lambda_k^2 + \lambda \tilde \gamma_k }.
\label{eq:DLcompl}
\end{equation}
Here, $\gamma_k$ is the relaxation time and $\omega_k$ is the characteristic frequency for the $k$-th term. Also, $\tilde{\gamma}_k = \gamma_k/c,\ \lambda_k = \omega_k/c$, and $\eta_k = 2\pi a f_k\omega_p^2/c\gamma_k$.  In Table  \ref{tab:DrDirLor} we show the parameters using the calculated values for graphite \cite{Djurisic:1999:Opog}.

\begin{table}[ht]
	\centerline{ \begin{tabular}{|c|c|c|c||c|c||}\hline\hline
			$\eta_k$ & &$\gamma_k$ & $eV$ & $\omega_k$ & $eV$ \\ \hline\hline
			$\eta_0$ & $0.01712$ &$\gamma_0$ & $6.365$ &- &-  \\ 
			$\eta_1$ & $0.13855$ &$\gamma_1$ & $4.102$ & $\omega_1$ & $0.275$  \\
			$\eta_2$ & $0.05949$ &$\gamma_2$ & $7.328$ & $\omega_2$ & $3.508$   \\
			$\eta_3$ & $0.37991$ &$\gamma_3$ & $1.414$ & $\omega_3$ & $4.451$  \\   
			$\eta_4$ & $0.08462$ &$\gamma_4$ & $0.46$ & $\omega_4$ & $13.591$ \\
			$\eta_5$ & $1.09548$ &$\gamma_5$ & $1.862$ & $\omega_5$ & $14.226$  \\
			$\eta_6$ & $0.30039$ &$\gamma_6$ & $11.922$& $\omega_6$ & $15.55$  \\
			$\eta_7$ & $0.03983$ &$\gamma_7$ & $39.091$& $\omega_7$ & $32.011$\\ \hline   
		\end{tabular}%
		\begin{tabular}{|c|c||c|c||}\hline\hline
			$\tilde\gamma_k$ & $\frac{1}{nm}$ & $\lambda_k$  & $\frac{1}{nm} $\\ \hline\hline
			$\tilde\gamma_0$ & $0.0322$ &- &-  \\ 
			$\tilde\gamma_1$ & $0.0207$ & $\lambda_1$ & $0.0014$  \\
			$\tilde\gamma_2$ & $0.0371$   & $\lambda_2$&$0.0177$   \\
			$\tilde\gamma_3$ & $0.0072$ & $\lambda_3$ & $0.0225$ \\   
			$\tilde\gamma_4$ & $0.0023$ & $\lambda_4$ & $0.0688$  \\
			$\tilde\gamma_5$ & $0.0094$ & $\lambda_5$ & $0.0721$  \\
			$\tilde\gamma_6$ & $0.0604$ & $\lambda_6$ & $0.0788$  \\
			$\tilde\gamma_7$ & $0.1981$ & $\lambda_7$ & $0.1622$  \\ \hline   
		\end{tabular}		
			}\caption{Parameters for the Drude-Lorentz model of a graphene sheet in 3D graphite}\label{tab:DrDirLor}
\end{table}

We note that the optical response in the infrared regime for 3D graphite and an isolated graphene is different. While $\sigma$ for graphite exhibits a Drude-like behavior, the graphene optical conductivity is constant. This difference is attributed to the different electronic structure characteristics for the two systems \cite{Marinopoulos:2004:isotoaawdrog}. To ensure that the $\sigma=const$ feature is captured, the single graphene conductivity $\widetilde{\eta}_{DL}$  is obtained by using a characteristic distance $a=0.224$ $nm$. In addition, we require that $\widetilde{\eta}_{DL}(0)$ coincides with $\eta_{gr}$ at zero frequency as:
\begin{equation}
\widetilde{\eta}_{DL} (\lambda) = \eta_{DL} (\lambda) \frac{\eta_{gr}}{\eta_0}.\label{eq:DLGr}
\end{equation} 

\subsection{Atomic Polarizabilities}\label{Sec:AppC} 
In general, the atomic polarizability can be represented as a multioscillator model in the following form
\begin{equation}
\alpha (\lambda) = \sum_{k=1}^{m}\frac{\alpha_k}{1 + \frac{\lambda^2}{\lambda_{a,k}^2}}, \label{eq:mosc}
\end{equation}
where the imaginary frequency axis is used. 

For lighter atoms, such as $\mathrm{H}_2$, He and He* one-oscillator model is typically used in Ref. \cite{Churkin:2010:CohaDmodibgaHHoNa}. Thus we utilize available data in  \cite{Rauber:1982:Sdibaam,Bruehl:2002:TvdWpbmaassNdevt} summarized as 
\begin{table}[ht]
	\centerline{ \begin{tabular}{||c|c|c||}\hline\hline
			Atom & $\alpha (0)(a.u.)$   & $\omega_a  (eV)$  \\ \hline\hline
			$\mathrm{H}$ & $4.5$ &$11.65$  \\ 
			$\mathrm{H}_2$ & $5.439$ &$14.09$  \\
			$\mathrm{He}$ & $1.384$ &$27.64$   \\
			$\mathrm{He}^*$ & $315.77$ &$1.18$  \\  \hline   
		\end{tabular}}\caption{Polarizability parameters of the single-oscillator model for several atoms (\ref{eq:mosc}). Here $1 \ a.u.  = 0.1482 \ \mathring{A}^3$. }\label{tab:Polar}
\end{table}

For heavier atoms, such as Na, K, Rb, Cs and Fe consistent data for the one-oscillator polarizability is not readily found. However, we use numerical results in \cite{Derevianko:1999:HCoDCSDPaAICfAA}, where precise calculations of the $\alpha(0)$ and their vdW coefficients $C_3^a$ and $C_6^a$ vdW coefficients are reported. Using the following relations,
\begin{equation}
C_3^a = \frac{1}{4\pi}\int_{0}^{\infty} d \omega \alpha (i\omega),\ C_6^a = \frac{3}{\pi}\int_{0}^{\infty} d \omega \alpha^2 (i\omega),\label{eq:c3c6}
\end{equation}
\begin{equation}
\alpha (i\omega)_{\omega =0} = \alpha (0), \ \lim_{\omega \to \infty} \omega^2 \alpha (i\omega) = N,\label{eq:0N}
\end{equation} 
it is realized that a two-oscillator model (4 parameters) for $\alpha(\omega)$ is needed in order to a have self-sustained solution. The results obtained are shown in Table \ref{tab:parameters}. 
\begin{table}[ht]
	\centerline{\begin{tabular}{||c||c|c|c|c|||c|c||}
			\hline & \multicolumn{4}{|c|||}{Two-oscillator} & \multicolumn{2}{c||}{Single-oscillator} \\ 
			\hline\hline  Atom & $\alpha_1 (a.u.)$ & $\omega_1(eV)$ & $\alpha_2 (a.u.)$ & $\omega_2(eV)$ & $\alpha_0 (a.u.)$ & $\omega_0(eV)$ \\ 
			\hline Na  & $162.1$ & $2.12$  & $0.547$ & $116.4$ & $162.6$  & $2.13$  \\ 
			\hline K & $288.4$ & $1.66$ & $1.754$ & $87.0$ & $290.2$ & $1.68$ \\ 
			\hline  Rb & $316.7$ & $1.65$ & $1.85$ & $119.6$ & $318.6$ & $1.68$ \\ 
			\hline  Cs & $397.3$ & $1.53$ & $2.597$ & $123.8$ & $399.9$ & $1.55$ \\ 
			\hline  Fe & $307.8$ & $1.75$ & $9.972$ & $42.8$ & $317.8$ & $1.89$ \\ 
			\hline 
		\end{tabular}}\caption{Parameters of two- and single-oscillator models for several atoms.}\label{tab:parameters}
	\end{table}

It is evident that the first oscillator gives the dominant contribution to the atomic polarizability. One further notes that if only $C_6^a$ coefficient is used together with $\alpha (i\omega)_{\omega =0} = \alpha (0)$, the obtained data for the atomic polarizability and characteristic frequency are very similar to the first oscillator parameters if obtained via the $C_6^a$ and $C_3^a$ two-oscillator scheme. 

The calculations for the Casimir-Polder force are not affected significantly by using the atomic polarizability via one- or two-oscillator representation if the regime of interest is at larger separations. For shorter separations, however, the difference can be significant. For example, the relative error for $C_3 (a,\cN)$ in Eq. (\ref{eq:C3}) is found to be on the order of  $15\%$ (Na), $23\%$ (K), $28\%$ (Rb), $33\%$ (Cs), $38\%$ (Fe). Therefore, for the calculations of the vdW coefficient in Eq. (\ref{eq:mosc}), the two oscillator model for the atomic polarzaibility is utilized.

\input{papcp.bbl.tex}
\end{document}

%% file: papcp.bbl.tex
%